\documentclass[a4paper,12pt]{article}
\usepackage[british]{babel}
\usepackage{amsfonts}
\usepackage{amsfonts}
\usepackage{amsmath}
\usepackage{amssymb}
\usepackage{graphicx}
\begin{document}

\begin{center}
\baselineskip=1.5
\normalbaselineskip{\Large Generalised cascades}

{\large S\'{\i}lvio M. Duarte Queir\'{o}s}\footnote{%
email address: Silvio.Queiros@unilever.com, sdqueiro@googlemail.com}

\baselineskip=1.0 \normalbaselineskip

\textit{Unilever R\&D Port Sunlight, Quarry Road East, Wirral, CH63 3JW UK \\%
[0pt]
}

\baselineskip=1.0 \normalbaselineskip

{\small (20th March 2009)}
\end{center}

\baselineskip=1.0 \normalbaselineskip

\subsection*{\protect\bigskip Abstract}

In this manuscript we give thought to the aftermath on the stable
probability density function when standard multiplicative cascades are
generalised cascades based on the $q$-product of Borges that emerged in the
context of non-extensive statistical mechanics.

\section{Introduction}

\label{intro}

In the twenty years that have elapsed since the publication of the
non-additive entropy $S_{q}$, also fairly known as \emph{Tsallis entropy}
\cite{ct-1988}, many applications and connections to natural and man-mind
phenomena have been established \cite{applications}. One of the most
exciting which that have emerged within the non-extensive scope is the
definition of a whole new set of mathematical operations/functions that goes
from the generalised algebra independently defined by Borges \cite{borges}
and Nivanen \textit{et al.} \cite{nivanen} and the integro-differential
operators by Borges to the $q$-trigonometric functions \cite{borges-tese}.
Besides its inherent beauty, these generalisations have found its own field
of applicability. Namely, the $q$-product plays a primary role in the
definition of the $q$-Fourier transform \cite{sabir}, thus in $q$-Central
Limit Theorem \cite{sabir1}, whereas the generalised trigonometric functions
have been quite successful in describing the critical behaviour of a class
of composed materials known as manganites \cite{manganites}. In this
article, we inquire into the possible applications of the $q$-product in the
generation of random variables and its consequence on the definition of a
new class of probability density functions.

\section{Preliminaries: the $q$-product}

\label{preliminar}

The $q$-product, $\otimes _{q}$, has been introduced with the purpose to
find a functional form that is able to generalise in a non-extensive way the
mathematical identity,
\begin{equation}
\exp \left[ \ln \,x+\ln \,y\right] =x\times y,\qquad \left( x,y>0\right) ,
\end{equation}%
so that the equality,
\begin{equation}
x\otimes _{q}y\equiv \exp _{q}\left[ \ln _{q}\,x+\ln _{q}\,y\right] ,
\label{q-product}
\end{equation}%
holds. The representations $\ln _{q}\left( .\right) $ and $\exp _{q}\left(
.\right) $ correspond to the $q$-logarithm \cite{ct-quimica},
\begin{equation}
\ln _{q}\left( x\right) \equiv \frac{x^{1-q}-1}{1-q},\qquad \left( x>0,q\in
\Re \right) ,
\end{equation}
and its inverse, the $q$-exponential,
\begin{equation}
\exp _{q}\left( x\right) \equiv \left[ 1+\left( 1-q\right) \,x\right] ^{%
\frac{1}{1-q}},\qquad \left( x,q\in \Re \right) ,
\end{equation}

respectively ($\exp _{q}\left( x\right) =0$ if $1+(1-q)\,x\leq 0$). For $%
q\rightarrow 1$, the equation (\ref{q-product}) recovers the usual property,
\begin{equation*}
\ln \left( x\times y\right) =\ln \,x+\ln \,y
\end{equation*}
($x,y>0$), with $x\times y\equiv x\otimes _{1}y$. Its inverse operation, the
$q$-division, $x\oslash _{q}y$, verifies the following equality $\left(
x\otimes _{q}y\right) \oslash _{q}y=x$.

Bearing in mind that the $q$-exponential is a non-negative function, the $q$%
-product must be restricted to the values of $x$ and $y$ that respect the
condition,
\begin{equation}
\left\vert x\right\vert ^{1-q}+\left\vert y\right\vert ^{1-q}-1\geq0.
\label{cond-q-prod}
\end{equation}
Moreover, we can extend the domain of the $q$-product to negative values of $%
x$ and $y$ writing it as,
\begin{equation}
x\otimes_{q}y\equiv\mathrm{\ sign}\left( x\,y\right) \exp_{q}\left[ \ln
_{q}\,\left\vert x\right\vert +\ln_{q}\,\left\vert y\right\vert \right] .
\label{q-product-new}
\end{equation}
Regarding some key properties of the $q$-product we mention:

\begin{enumerate}
\item $x\otimes_{1}y=x\ y$;

\item $x\otimes_{q}y=y\otimes_{q}x$;

\item $\left( x\otimes_{q}y\right) \otimes_{q}z=x\otimes_{q}\left(
y\otimes_{q}z\right) =\left[ x^{1-q}+y^{1-q}-2\right] ^{\frac{1}{1-q}}$;

\item $\left( x\otimes_{q}1\right) =x$;

\item $\ln_{q}\left[ x\otimes_{q}y\right] \equiv\ln_{q}\,x+\ln_{q}\,y$;

\item $\ln_{q}\left( x\,y\right) =\ln_{q}\left( x\right) +\ln_{q}\left(
y\right) +\left( 1-q\right) \ln_{q}\left( x\right) \ln_{q}\left( y\right) $;

\item $\left( x\otimes_{q}y\right) ^{-1}=x^{-1}\otimes_{2-q}y^{-1}$;

\item $\left( x\otimes _{q}0\right) =\left\{
\begin{array}{ccc}
0 &  & \mathrm{if\ }\left( q\geq 1\ \mathrm{and\ }x\geq 0\right) \ \mathrm{%
or\ if\ }\left( q<1\ \mathrm{and\ }0\leq x\leq 1\right)  \\
&  &  \\
\left( x^{1-q}-1\right) ^{\frac{1}{1-q}} &  & \mathrm{otherwise}%
\end{array}%
\right. $
\end{enumerate}

For particular values of $q$, \textit{e.g.}, $q=1/2$, the $q$-product
provides nonnegative values at points for which the inequality $\left\vert
x\right\vert ^{1-q}+\left\vert y\right\vert ^{1-q}-1<0$ is verified.
According to the cut-off of the $q$-exponential, a value of zero for $%
x\otimes _{q}y$ is set down in these cases. Restraining our analysis of the
Eq. (\ref{cond-q-prod}) to the sub-space $x,y>0$, we can observe that for $%
q\rightarrow -\infty $ the region $\left\{ 0\leq x\leq 1,0\leq y\leq
1\right\} $ is not defined. As the value of $q$ increases, the forbidden
region decreases its area, and when $q=0$, we have the limiting line given
by $x+y=1$, for which $x\otimes _{0}y=0$. Only for $q=1$, the entire set of $%
x$ and $y$ real values of has a defined value for the $q$-product. For $q>1$%
,\ the condition (\ref{cond-q-prod}) implies a region, $\left\vert
x\right\vert ^{1-q}+\left\vert y\right\vert ^{1-q}=1$ for which the $q$%
-product diverges. This undefined region augments its area as $q$ goes to
infinity. When $q=\infty $, the $q$-product is only defined in $\left\{
x\geq 0,0\leq y\leq 1\right\} \cup \left\{ 0\leq x\leq 1,y>1\right\} $.
Illustrative plots are presented in Fig.\ (1) of Ref. \cite{part1}.

\section{Multiplicative processes as generators of distributions}

\label{multiplicative}

Multiplicative processes, particularly stochastic multiplicative processes,
have been the source of plenty of models applied in several fields of
science and knowledge. In this context, we can name the study of fluid
turbulence \cite{turbulence}, fractals \cite{feder}, finance \cite%
{mandelbrot}, linguistics \cite{murilinho}, etc. Specifically,
multiplicative processes play a very important role on the emergence of the
log-Normal distribution as a natural and ubiquitous distribution. In simple
terms, the log-Normal distribution is the distribution of a random variable
whose logarithm is associated with a Normal distribution \cite%
{log-normal-book},%
\begin{equation}
p\left( x\right) =\frac{1}{\sqrt{2\pi }\sigma x}\exp \left[ -\frac{\left(
\ln x-\mu \right) ^{2}}{2\sigma ^{2}}\right] .  \label{log-normal}
\end{equation}%
With regard to the dynamical origins of the log-Normal distribution, several
processes have been thought up to generate it. In this work we highlight the
two most famous --- the \emph{law of proportionate effect}~\cite{gibrat},
the \emph{theory of breakage}~\cite{kolmogorov} or from Langevin-like
processes \cite{fa}. We shall now give a brief view of the former; Let us
consider a variable $\tilde{Z}$ obtained from a multiplicative random
process,%
\begin{equation}
\tilde{Z}=\prod\limits_{i=1}^{N}\tilde{\zeta}_{i},  \label{product}
\end{equation}%
where $\tilde{\zeta}_{i}$ are nonnegative microscopic variables associated
with a distribution $f^{\prime }\left( \tilde{\zeta}\right) $. If we
consider the following transform of variables $Z\equiv \ln \tilde{Z}$, then
we have,%
\begin{equation*}
Z=\sum\limits_{i=1}^{N}\zeta _{i},
\end{equation*}%
with $\zeta \equiv \ln \tilde{\zeta}$. Assume now $\zeta $ as a variable
associated with a distribution $f\left( \zeta \right) $ with average $\mu $
and variance $\sigma ^{2}$. Then, $Z$ converges to the Gaussian distribution
in the limit of $N$ going to infinity as entailed by the Central Limit
Theorem \cite{araujo}. Explicitly, considering that the variables $\zeta $
are independently and identically distributed, the Fourier Transform of $%
p\left( Z^{\prime }\right) $ is given by,

\begin{equation}
\mathcal{F}\left[ p\left( Z^{\prime }\right) \right] \left( k\right) =\left[
\int_{-\infty }^{+\infty }e^{i\,k\,\frac{\zeta }{N}}\,f\left( \zeta \right)
\,d\zeta \right] ^{N},  \label{fourier1}
\end{equation}%
where $Z^{\prime }=N^{-1}Z$. For all $N$, the integrand can be expanded as,%
\begin{equation}
\begin{array}{c}
\mathcal{F}\left[ p\left( Z^{\prime }\right) \right] \left( k\right) =\left[
\sum\limits_{n=0}^{\infty }\frac{\left( ik\right) ^{n}}{n!}\frac{%
\left\langle \zeta ^{n}\right\rangle }{N}\right] ^{N}, \\
\\
\mathcal{F}\left[ p\left( Z^{\prime }\right) \right] \left( k\right) =\exp
\left\{ N\ln \left[ 1+ik\frac{\left\langle \zeta \right\rangle }{N}-\frac{1}{%
2}k^{2}\frac{\left\langle \zeta ^{2}\right\rangle }{N^{2}}+O\left(
N^{-3}\right) \right] \right\} ,%
\end{array}%
\end{equation}%
expanding the logarithm,%
\begin{equation}
\mathcal{F}\left[ P\left( Z^{\prime }\right) \right] \left( k\right) \approx
\exp \left[ ik\mu -\frac{1}{2N}k^{2}\sigma ^{2}\right] .
\end{equation}%
Applying the inverse Fourier Transform, and reverting the $Z^{\prime }$
change of variables we finally obtain,%
\begin{equation}
p\left( Z\right) =\frac{1}{\sqrt{2\,\pi \,N}\sigma }\exp \left[ -\frac{%
\left( Z-N\,\mu \right) ^{2}}{2\,\sigma ^{2}\,N}\right] .
\end{equation}%
We can define the attracting distribution in terms of the original
multiplicative random process, yielding the log-Normal distribution \cite%
{log-normal-book},%
\begin{equation}
p\left( \bar{Z} \right) =\frac{1}{\sqrt{2\,\pi \,N}\sigma \,\bar{Z}}\exp \left[ -%
\frac{\left( \ln \bar{Z}-N\,\mu \right) ^{2}}{2\,\sigma ^{2}\,N}\right] .
\end{equation}

\begin{figure}[tbh]
\begin{center}
\includegraphics[width=0.65\columnwidth,angle=0]{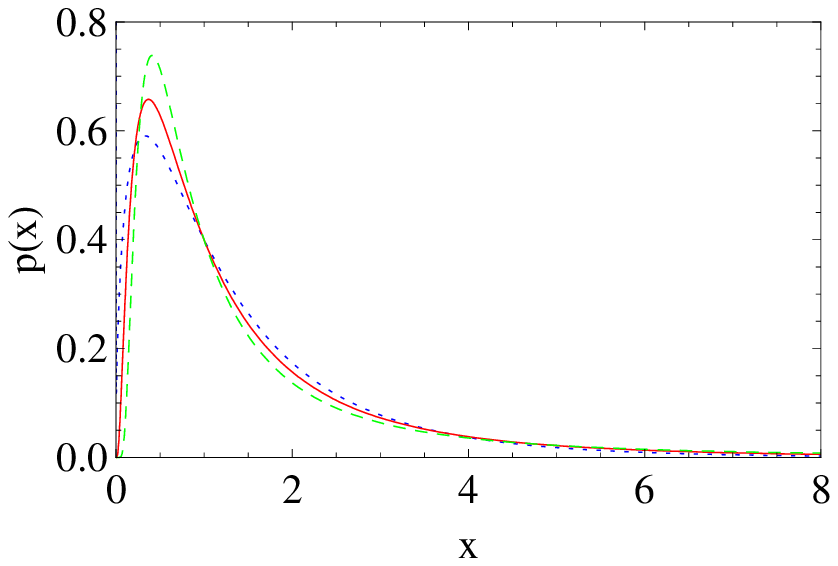} %
\includegraphics[width=0.65\columnwidth,angle=0]{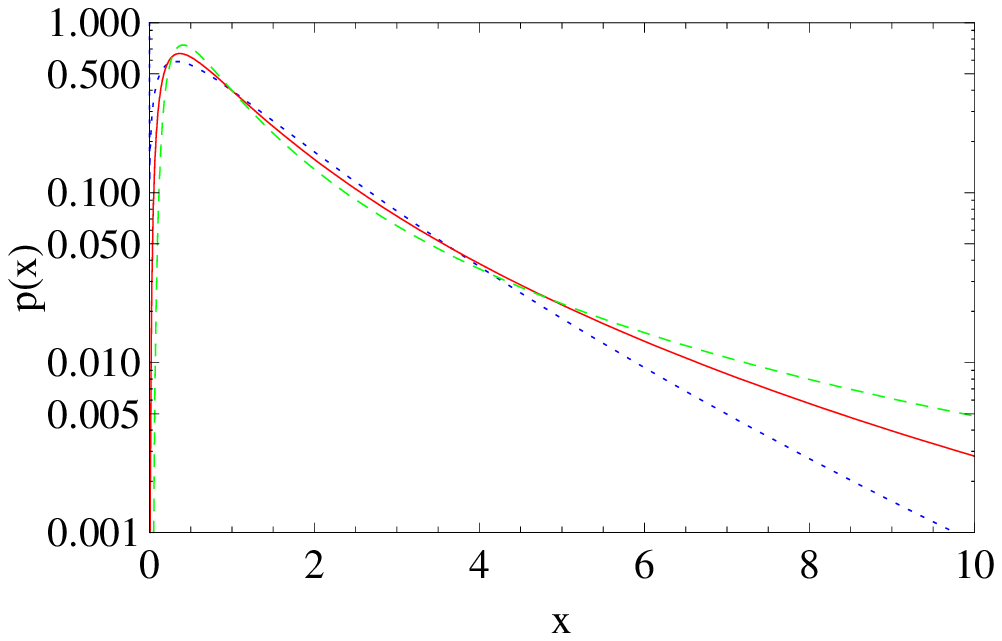} %
\includegraphics[width=0.65\columnwidth,angle=0]{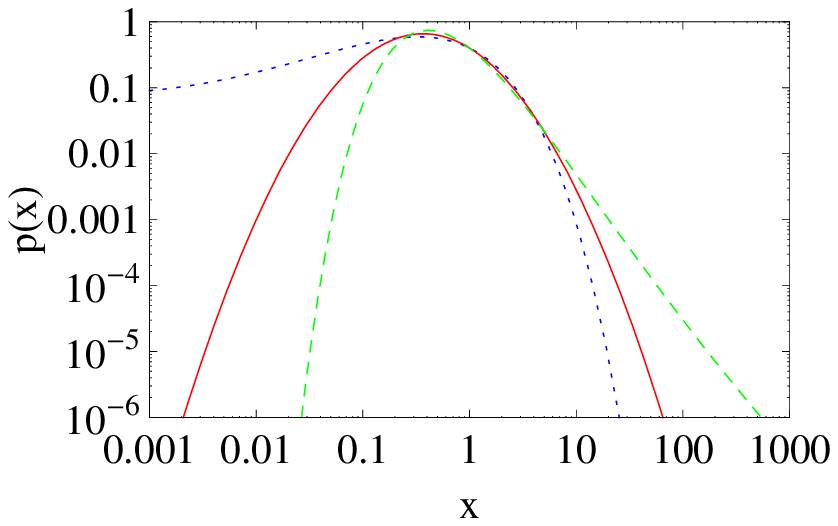}
\end{center}
\caption{Plots of the Eq. (\protect\ref{qlog-normal}) \textit{vs} $x$ for $q=%
\frac {4}{5}$ (dotted line), $q=1$ (full line) and $q=\frac{5}{4}$ (dashed
line) in linear-linear scale (upper), log-linear (centre), log-log (lower).}
\label{fig-pdf}
\end{figure}

Although this distribution with two parameters, $\mu $ and $\sigma $, is
able to appropriately describe a large variety of data sets, there are cases
for which the log-Normal distribution fails statistical testing \cite%
{log-normal-book}. In some of these cases, such a failure has been overcome
by introducing different statistical distributions (e.g., Weibull
distributions) or changing the 2-parameter log-Normal distribution by a
3-parameter log-Normal distribution,%
\begin{equation}
p\left( x\right) =\frac{1}{\sqrt{2\,\pi }\sigma \,\left( x-\theta \right) }%
\exp \left[ -\frac{\left( \ln \left[ x-\theta \right] -\mu \right) ^{2}}{%
2\,\sigma ^{2}}\right] .
\end{equation}%
In the sequel of this work we present an alternative procedure to generalise
the Eq. (\ref{log-normal}). The motivation for this proposal comes from
changing the $N$ products in Eq. (\ref{product}) by $N$ $q$-products,%
\begin{equation}
\tilde{Z}=\underset{}{\prod\limits_{i=1}^{N}}^{(q)}\tilde{\zeta}_{i}\equiv
\tilde{\zeta}_{1}\otimes _{q}\tilde{\zeta}_{2}\otimes _{q}\ldots \otimes _{q}%
\tilde{\zeta}_{N}.  \label{p-product}
\end{equation}%
Applying the $q$-logarithm we have a sum of $N$ terms. If every term is
identically and independently distributed, then for variables $\zeta
_{i}=\ln _{q}$ $\tilde{\zeta}_{i}$ with finite variables we have a Gaussian
has stable distribution, \textit{i.e.}, a Gaussian distribution in the $q$%
-logarithm variable. From this scenario we can obtain our $q$\emph{-log
Normal probability density function},%
\begin{equation}
p\left( x\right) =\frac{1}{\mathcal{Z}_{q}\,x^{q}}\exp \left[ -\frac{\left(
\ln _{q}\,x-\mu \right) ^{2}}{2\,\sigma ^{2}}\right] ,\qquad \left( x\geq
0\right) ,  \label{qlog-normal}
\end{equation}%
with the normalisation,%
\begin{equation}
\mathcal{Z}_{q}=\left\{
\begin{array}{ccc}
\sqrt{\frac{\pi }{2}}\mathrm{erfc}\left[ -\frac{1}{\sqrt{2}\sigma }\left(
\frac{1}{1-q}+\mu \right) \right] \sigma  & if & q<1 \\
&  &  \\
\sqrt{\frac{\pi }{2}}\mathrm{erfc}\left[ \frac{1}{\sqrt{2}\sigma }\left(
\frac{1}{1-q}+\mu \right) \right] \sigma  & if & q>1.%
\end{array}%
\right.
\end{equation}%
In the limit of $q$ equal to $1$, $\ln _{q\rightarrow 1}x=\ln x$ and $%
\mathcal{Z}_{q\rightarrow 1}=\sqrt{2\,\pi }\sigma $ and the usual log-Normal
is recovered thereof (erfc stands for complementary error function). Typical
plots for cases with $q=\frac{4}{5}$, $q=1$, $q=\frac{5}{4}$ are depicted in
Fig.~\ref{fig-pdf}.

The raw statistical moments,%
\begin{equation}
\left\langle x^{n}\right\rangle \equiv \int_{0}^{\infty }x^{n}\,p\left(
x\right) \,dx,  \label{momentos}
\end{equation}%
can be analytically computed for $q<1$ giving \cite{gradshteyn},%
\begin{equation}
\left\langle x^{n}\right\rangle =\frac{\Gamma \left[ \nu \right] \exp \left[
-\frac{\gamma ^{2}}{8\,\beta }\right] D_{-\nu }\left[ \frac{\gamma }{\sqrt{%
2\,\beta }}\right] }{\sqrt{\beta ^{\nu }\,\pi }\sigma \left( 1-q\right)
\mathrm{erfc}\left[ -\frac{1}{\sqrt{2}\sigma }\left( \frac{1}{1-q}+\mu
\right) \right] },  \label{raw moment}
\end{equation}%
with%
\begin{equation}
\beta =\frac{1}{2\sigma ^{2}\left( 1-q\right) ^{2}};\quad \gamma =-\frac{%
1+\mu \,\left( 1-q\right) }{\left( 1-q\right) ^{2}\,\sigma ^{2}};\quad \nu
=1+\frac{n}{1-q},
\end{equation}%
where $D_{-a}\left[ z\right] $ is the \emph{parabolic cylinder function}
\cite{paraboliccylinderd}. For $q>1$, the raw moments are given by an
expression quite similar to the Eq. (\ref{raw moment}) with the argument of
the erfc replaced by $\frac{1}{\sqrt{2}\sigma }\left( \frac{1}{1-q}+\mu
\right) $. However, the finiteness of the raw moments is not guaranteed for
every $q>1$ for two very related reasons. First, according to the definition
of $D_{-\nu }\left[ z\right] $, $\nu $ must be greater than $0$. Second, the
core of the probability density function, $\exp \left[ -\frac{\left( \ln
_{q}\,x-\mu \right) ^{2}}{2\,\sigma ^{2}}\right] $, does not vanish in the
limit of $x$ going to infinity $\infty $,%
\begin{equation}
\lim_{x\rightarrow \infty }\exp \left[ -\frac{\left( \ln _{q}\,x-\mu \right)
^{2}}{2\,\sigma ^{2}}\right] =\exp \left[ -\frac{\gamma ^{2}}{2}\right] .
\end{equation}%
This means that the limit $p\left( x\rightarrow \infty \right) =0$ is
introduced by the normalisation factor $x^{-q}$, which comes from redefining
the Gaussian of variables,
\begin{equation}
y\equiv \ln _{q}x,
\end{equation}%
as a distribution of variables $x$. Because of that, if the moment surpasses
the value of $q$, then the integral (\ref{momentos}) diverges.

\section{Examples of cascade generators}

\label{examplo}

In this section, we discuss the upshot of two simple cases in which the
dynamical process described in the previous section is applied. We are going
to verify that the value of $q$ influences the nature of the attractor in
probability space.

\subsection{Compact distribution $[0,b]$}

\begin{figure}[tbh]
\begin{center}
\includegraphics[width=0.65\columnwidth,angle=0]{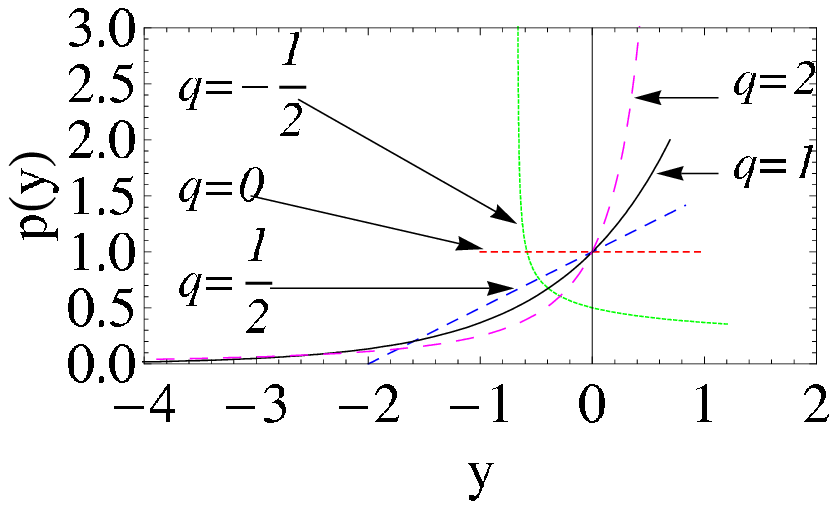}
\end{center}
\caption{Plots of the Eq. (\protect\ref{q-uniform}) \textit{vs} $y$ for $b=2$
and the values of $q$ presented in the text. }
\label{fig-flat}
\end{figure}

Let us consider a compact distribution for indentically and independently
distributed variables $x$ within the interval $0$ and $b$. Following what we
have described in the preceding section, we can transform our generalised\
multiplicative process into a simple additive process of $y_{i}$ variables
which are now distributed in conformity with the distribution,%
\begin{equation}
p^{\prime}\left( y\right) =\frac{1}{b}\left[ 1+\left( 1-q\right) y\right] ^{%
\frac{q}{1-q}},  \label{q-uniform}
\end{equation}
with $y$ defined between $\frac{1}{q-1}$ and $\frac{b^{1-q}-1}{1-q}$ if $q<1$%
, whereas $y$ ranges over the interval between $-\infty$ and $\frac{b^{1-q}-1%
}{1-q}$ when $q>1$. Some curves for the special case $b=2$ are plotted in
Fig.~\ref{fig-flat}.

\begin{figure}[tbh]
\begin{center}
\includegraphics[width=0.65\columnwidth,angle=0]{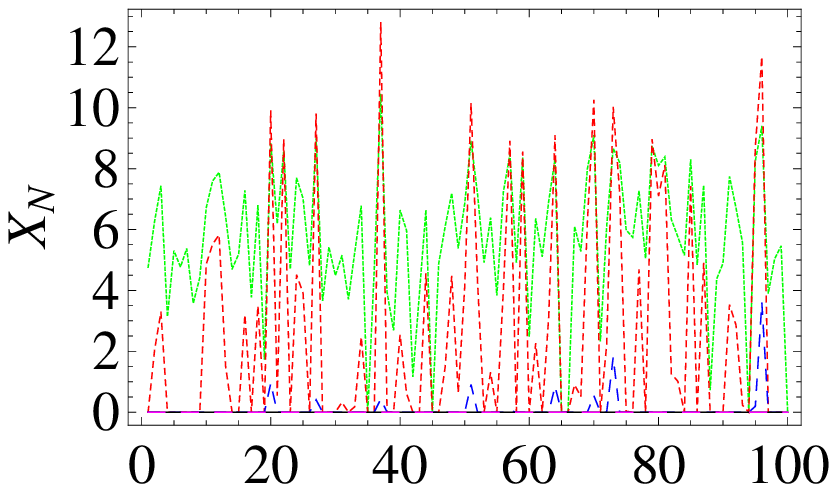} %
\includegraphics[width=0.65\columnwidth,angle=0]{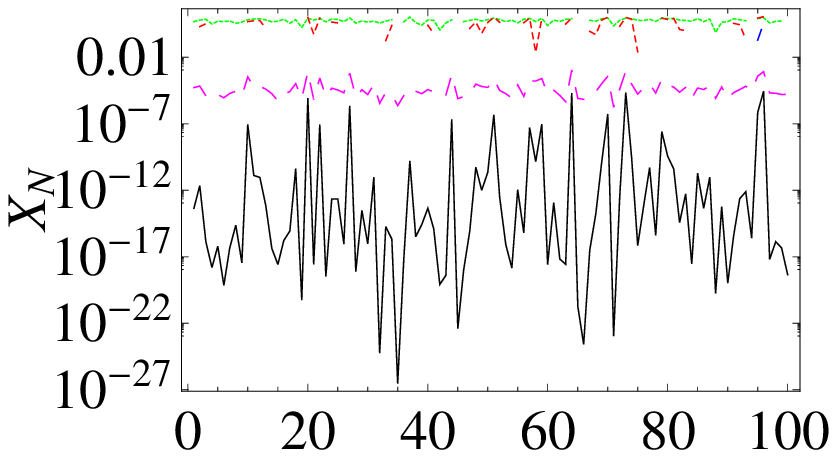}
\end{center}
\caption{Sets of random variables generated from the process (\protect\ref%
{p-product}) with $N=100$ and $q=-\frac{1}{2}$ (green), $0$ (red), $\frac{1}{%
2}$ (blue), $1$ (black), $\frac{5}{4}$ (magenta) in linear (upper panel) and
log scales (lower panel). The generating variable is uniformly distributed
within the interval $\left[ 0,1\right] $ as is the same for all of the cases
that we present. As visible, the value of $q$ deeply affects the values of $%
X_{N}=\tilde{Z}$. }
\label{fig-gen}
\end{figure}

If we look at the variance of this independent variable,%
\begin{equation}
\sigma _{y}^{2}=\left\langle y^{2}\right\rangle -\left\langle \mu
_{y}\right\rangle ^{2},
\end{equation}%
which is the moment whose finitude plays the leading role in the Central
Limit Theory, we verify that for $q>\frac{3}{2}$, we obtain a divergent
value,%
\begin{equation}
\sigma _{y}^{2}=\frac{b^{2-2\,q}}{\left( 3-2\,q\right) \left( 2-q\right) ^{2}%
}.
\end{equation}%
Hence, if $q<\frac{3}{2}$, we can apply the Lyapunov's central Limit theorem
and our attractor in the probability space is the Gaussian distribution. On
the other hand, if $q>\frac{3}{2}$, the L\'{e}vy-Gnedenko's version of the
central limit theorem \cite{levy}\ asserts that the attracting distribution
is a L\'{e}vy distribution with a tail exponent,
\begin{equation}
\alpha =\frac{1}{q-1}.
\end{equation}%
Furthermore, it is simple to verify that the interval $\left( \frac{3}{2}%
,\infty \right) $ of $q$ values\ maps onto the interval $\left( 0,2\right) $
of $\alpha $ values, which is precisely the interval of validity of the L%
\'{e}vy class of distributions that is defined by its Fourier Transform,%
\begin{equation}
\mathcal{F}\left[ L_{\alpha }\left( Y\right) \right] \left( k\right) =\exp %
\left[ -a\,\left\vert k\right\vert ^{\alpha }\right] .  \label{eq-levy}
\end{equation}%
In Fig.~\ref{fig-gen} we depict some sets generated by this process for
different values of $q$.

\subsection{$q$-log Normal distribution}

In this example, we consider the case of generalised multiplicative
processes in which the variables follow a $q$-log Normal distribution. In
agreement with what we have referred to in Sec. \ref{multiplicative}, the
outcome strongly depends on the value of $q$. Consequently, in the
associated $x$ space, if we apply the generalised process to $N$ variables $%
y=\ln _{q}\,x$ ($x\in \left[ 0,\infty \right) $) which follow a
Gaussian-like functional\footnote{%
Strictly speaking, we cannot use the term Gaussian distribution because it
is not defined in the interval $\left( -\infty ,\infty \right) $. The
limitations in the domain do affect the Fourier transform and thus the
result of the convolution of the probability density function.} form with
average $\mu $ and finite standard deviation $\sigma $, \textit{i.e.}, $%
\forall _{q<1} $ or $q>3$ in Eq.(\ref{qlog-normal}), the resulting
distribution in the limit of $N$ going to infinity corresponds to the
probability density function (\ref{qlog-normal}) with $\mu \rightarrow
N\,\mu $ and $\sigma ^{2}\rightarrow N\,\sigma ^{2}$. In respect of the
conditions of $q$ we have just mentioned here above, the $q$-log normal can
be seen as an asymptotic attractor, a stable attractor for $q=1$, and an
unstable distribution for the remaining cases with the resulting attracting
distribution being computed by applying the convolution operation.

\section{Final remarks}

In this manuscript we have introduced a modification in the multiplicative
process that has enabled us to present a modification on the log-Normal
distribution as well as other distributions with slow decay. This
distribution is controlled by an extra-parameter, $q$, when it is compared
with the regular 2-parameter log-Normal distribution, that can be
dynamically related to a change in the multiplicative random process.
Besides, it provides interesting mechanisms of on-off dynamics.

Regarding further applications, it is known that the standard log-normal
distribution is unfitted for several data sets. This 3-parameter log-Normal
probability function is expected to provide a better approach to these data
\cite{stats}.

\bigskip

{\small SMDQ thanks his colleagues at Unilever for discussions and financial
support of the Marie Curie Fellowship programme (European Union).}

\end{document}